\documentclass[twocolumn,showpacs,preprintnumbers]{revtex4}
\usepackage{amsmath}
\usepackage{amssymb}
\usepackage{amsfonts}
\usepackage{graphicx}
\usepackage{dcolumn}
\usepackage{bm}

\input{tcilatex}

\begin{document}

\title{Introduction to the study of entropy \\
in Quantum Games}
\author{Esteban Guevara Hidalgo$^{\dag \ddag }$}
\affiliation{$^{\dag }$Departamento de F\'{\i}sica, Escuela Polit\'{e}cnica Nacional,
Quito, Ecuador\\
$^{\ddag }$SI\'{O}N, Autopista General Rumi\~{n}ahui, Urbanizaci\'{o}n Ed%
\'{e}n del Valle, Sector 5, Calle 1 y Calle A \# 79, Quito, Ecuador}

\begin{abstract}
The present work is an introductory study about entropy its properties and
its role in quantum information theory. In a next work, we will use these
results to the analysis of a quantum game described by a density operator $%
\rho $ and with its entropy equal to von Neumann's.
\end{abstract}

\maketitle
\email{esteban\_guevarah@yahoo.es}

\section{Introduction}

In a recent work \cite{1} we proposed quantization relationships which would
let us describe and solution problems originated by conflicting or
cooperative behaviors among the members of a system from the point of view
of quantum mechanical interactions. Through these relationships we could
described a system through a density operator and its entropy would be given
by the von Neumann entropy. The quantum version of the replicator dynamics
is the equation of evolution of mixed states from quantum statistical
mechanics.

Since Shannon \cite{2}, information theory or the mathematical theory of
communication changed from an engineering discipline that dealed with
communication channels and codes \cite{3} to a physical theory \cite{4} in
where the introduction of the concepts of entropy and information were
indispensable to our understanding of the physics of measurement. Classical
information theory has two primary goals \cite{5}: The first is the
development of the fundamental theoretical limits on the achievable
performance when communicating a given information source over a given
communications channel using coding schemes from within a prescribed class.
The second goal is the development of coding schemes that provide
performance that is reasonably good in comparison with the optimal
performance given by the theory.

Quantum information theory may be defined \cite{6} as the study of the
achievable limits to information processing possible within quantum
mechanics. Thus, the field of quantum information has two tasks: First, it
aims to determine limits on the class of information processing tasks which
are possible in quantum mechanics and provide constructive means for
achieving information processing tasks. Quantum information theory appears
to be the basis for a proper understanding of the emerging fields of quantum
computation \cite{7,8}, quantum communication \cite{9,10}, and quantum
cryptography \cite{11,12}. Entropy is the central concept of information
theories. The quantum analogue of entropy appeared 21 years before Shannon's
entropy and generalizes Boltzmann's expression. It was introduced in quantum
mechanics by von Neumann \cite{13,14}. Entropy\textbf{\ }in quantum
information theory plays prominent roles in many contexts, e.g., in studies
of the classical capacity of a quantum channel \cite{15,16} and the
compressibility of a quantum source \cite{17,18}.

\section{Shannon, Entropy and Classical Information Theory}

Entropy \cite{2,19} is the central concept of information theory. In
classical physics, information processing and communication is best
described by Shannon information theory.

The Shannon entropy expresses the average information we expect to gain on
performing a probabilistic experiment of a random variable $A$ which takes
the value $a_{i}$ with the respective probability $p_{i}$. It also can be
seen as a measure of uncertainty before we learn the value of $A$. We define
the Shannon entropy of a random variable $A$ by%
\begin{equation}
H(A)\equiv H(p_{1},...,p_{n})\equiv -\sum_{i=1}^{n}p_{i}\log _{2}p_{i}\text{.%
}  \label{1}
\end{equation}%
The entropy of a random variable is completely determined by the
probabilities of the different possible values that the random variable
takes. Due to the fact that $p=(p_{1},...,p_{n})$ is a probability
distribution, it must satisfy $\sum_{i=1}^{n}p_{i}=1$ and $0\leq
p_{1},...,p_{n}\leq 1$. The Shannon entropy of the probability distribution
associated with the source gives the minimal number of bits that are needed
in order to store the information produced by a source, in the sense that
the produced string can later be recovered. Shannon formalized the
requirements for an information measure $H(p_{1},...,p_{n})$ with the
following criteria:

\begin{enumerate}
\item $H$ should be continuous in the $p_{i}$.

\item If the $p_{i}$ are all equal, i.e. $p_{i}=1/n$, then $H$ should be a
monotonic increasing function of $n$.

\item $H$ should be objective: If a choice be broken down into two
successive choices, the original $H$ should be the weighted sum of the
individual values of $H$.%
\begin{eqnarray}
H(p_{1},...,p_{n}) &=&H(p_{1}+p_{2},p_{3,}...,p_{n})  \notag \\
&&+(p_{1}+p_{2})H(\frac{p_{1}}{p_{1}+p_{2}},\frac{p_{2}}{p_{1}+p_{2}})\text{.%
}  \label{2}
\end{eqnarray}
\end{enumerate}

Suppose $A$ and $B$ are two random variables. The \textit{joint entropy} $%
H(A,B)$ measures our total uncertainty about the pair $(A,B)$. The joint
entropy $H(A,B)$ is defined by%
\begin{equation}
H(A,B)\equiv -\sum_{i,j}p_{ij}\log _{2}p_{ij}  \label{3}
\end{equation}%
while%
\begin{gather}
H(A)=-\sum_{i,j}p_{ij}\log _{2}\sum_{j}p_{ij}\text{,}  \label{4} \\
H(B)=-\sum_{i,j}p_{ij}\log _{2}\sum_{i}p_{ij}\text{,}  \label{5}
\end{gather}%
where $p_{ij}$ is the joint probability to find $A$ in state $a_{i}$ and $B$
in state $b_{j}$.

The \textit{conditional entropy} $H(A\mid B)$ is a measure of how uncertain
we are about the value of $A$, given that we know the value of $B$. The
entropy of $A$ conditional on knowing that $B$ takes the value $b_{j}$ is
defined by%
\begin{gather}
H(A\mid B)\equiv H(A,B)-H(B)\text{,}  \notag \\
H(A\mid B)=-\sum_{i,j}p_{ij}\log _{2}p_{i\mid j}\text{,}  \label{6}
\end{gather}%
where $p_{i\mid j}=\frac{p_{ij}}{\sum_{i}p_{ij}}$ is the conditional
probability that $A$ is in state $a_{i}$ given that $B$ is in state $b_{j}$.

The \textit{mutual or correlation entropy} $H(A:B)$ of $A$ and $B$ measures
how much information $A$ and $B$ have in common. The mutual or correlation
entropy $H(A:B)$ is defined by%
\begin{gather}
H(A:B)\equiv H(A)+H(B)-H(A,B)\text{,}  \notag \\
H(A:B)\equiv -\sum_{i,j}p_{ij}\log _{2}p_{i:j}\text{,}  \label{7}
\end{gather}%
where $p_{i:j}$ is the mutual probability defined as $p_{i:j}=\frac{%
\sum_{i}p_{ij}\sum_{j}p_{ij}}{p_{ij}}$. The mutual or correlation entropy
also can be expressed through the conditional entropy via%
\begin{gather}
H(A:B)=H(A)-H(A\mid B)\text{,}  \label{8} \\
H(A:B)=H(B)-H(B\mid A)\text{.}  \label{9}
\end{gather}%
The joint entropy would equal the sum of each of $A$'s and $B$'s entropies
only in the case that there are no correlations between $A$'s and $B$'s
states. In that case, the mutual entropy or information vanishes and we
could not make any predictions about $A$ just from knowing something about $%
B $ \cite{20}.

The \textit{relative entropy} $H(p\parallel q)$ measures the closeness of
two probability distributions, $p$ and $q$, defined over the same random
variable $A$. We define the relative entropy of $p$ with respect to $q$ by%
\begin{gather}
H(p\parallel q)\equiv \sum_{i}p_{i}\log _{2}p_{i}-\sum_{i}p_{i}\log _{2}q_{i}%
\text{,}  \notag \\
H(p\parallel q)\equiv -H(A)-\sum_{i}p_{i}\log _{2}q_{i}\text{.}  \label{10}
\end{gather}%
The relative entropy is non-negative, $H(p\parallel q)\geq 0$, with equality
if and only if $p=q$. The classical relative entropy of two probability
distributions is related to the probability of distinguishing the two
distributions after a large but finite number of independent samples
(Sanov's theorem) \cite{3}.

Lets review some basic properties of entropy \cite{3}:

\begin{enumerate}
\item $H(A,B)=H(B,A),H(A:B)=H(B:A)$.

\item $H(B\mid A)\geq 0$ and thus $H(A:B)\leq H(B)$, with equality if and
only if $B=f(A)$.

\item $H(A)\leq H(A,B)$, with equality if and only if $B=f(A).$

\item Subadditivity: $H(A,B)\leq H(A)+H(B)$ with equality if and only if $A$
and $B$ are independent random variables.

\item $H(B\mid A)\leq H(B)$ and thus $H(A:B)\geq 0$, with equality in each
if and only if $A$ and $B$ are independent random variables.

\item Strong subadditivity: $H(A,B,C)+H(B)\leq H(A,B)+H(B,C)$.
\end{enumerate}

Conditioning reduces entropy%
\begin{equation}
H(A\mid B,C)\leq H(A\mid B)  \label{11}
\end{equation}%
and for a set of random variables $A_{1},...,A_{n}$ and $B$, the chaining
for conditional entropies is%
\begin{equation}
H(A_{1},...,A_{n}\mid B)=\tsum_{i=1}^{n}H(A_{i}\mid B,A_{1},...,A_{i-1})%
\text{.}  \label{12}
\end{equation}%
Suppose $A\rightarrow B\rightarrow C$ is a Markov chain. Then%
\begin{gather}
H(A)\geq H(A:B)\geq H(A:C)\text{,}  \label{13} \\
H(C:B)\geq H(C:A)\text{.}  \label{14}
\end{gather}%
The first inequality (\ref{13}) is saturated if and only if, given $B$, it
is possible to reconstruct $A$. The \textit{data processing inequality} (\ref%
{13}) states that the information we have available about a source of
information can only decrease with the time: once information has been lost,
it is gone forever. If a random variable $A$ is subject to noise, producing $%
B$, the data processing cannot be used to increase the amount of mutual
information between the output of the process and the original information $%
A $. The \textit{data pipelining inequality} (\ref{14}) says that any
information $C$ shares with $A$ must be information which $C$ also shares
with $B$; the information is \textquotedblleft pipelined\textquotedblright\
from $A$ through $B$ to $C$ \cite{6}.

\section{Von Neumann, Entropy and Quantum Information Theory}

Von Neumann \cite{13,14} defined the entropy of a quantum state $\rho $ by
the formula%
\begin{equation}
S(\rho )\equiv -Tr(\rho \ln \rho )  \label{15}
\end{equation}%
which is the quantum analogue of the Shannon entropy $H$ \cite{19}. The
entropy $S(\rho )$ is non-negative and takes its maximum value $\ln n$ when $%
\rho $ is maximally mixed, and its minimum value zero if $\rho $ is pure. If 
$\lambda _{i}$ are the eigenvalues of $\rho $ then von Neumann's definition
can be expressed as%
\begin{equation}
S(\rho )=-\tsum_{i}\lambda _{i}\ln \lambda _{i}\text{.}  \label{16}
\end{equation}%
The von Neumann entropy reduces to a Shannon entropy if $\rho $ is a mixed
state composed of orthogonal quantum states \cite{20}. If $U$ is a unitary
transformation, then%
\begin{equation}
S(\rho )=S(U\rho U^{\dagger })\text{.}  \label{17}
\end{equation}%
If a composite system $AB$ is in a pure state, then $S(A)=S(B)$. Suppose $%
\rho =\tsum_{i}p_{i}\rho _{i}$ where $p_{i}$ are probabilities, and $\rho
_{i}$ are density operators. Then%
\begin{equation}
S(\rho )\leq H(p_{i})+\tsum_{i}p_{i}S(\rho _{i})  \label{18}
\end{equation}%
with equality if and only if the states $\rho _{i}$ have support on
orthogonal subspaces, i.e. suppose $\left\vert i\right\rangle $ are
orthogonal states for a system $A$, and $\rho _{i}$ is any set of density
operators for another system $B$. Then%
\begin{equation}
S(\tsum_{i}p_{i}\left\vert i\right\rangle \left\langle j\right\vert \otimes
\rho _{i})=H(p_{i})+\tsum_{i}p_{i}S(\rho _{i})\text{,}  \label{19}
\end{equation}%
where $H(p_{i})$ is the Shannon entropy of the distribution $p_{i}$. The
entropy is a concave function of its inputs. That is, given real numbers $%
p_{i}$, satisfying $p_{i}\geq 0$, $\tsum_{i}p_{i}=1$, and its corresponding
density operators $\rho _{i}$, the entropy satisfies the equation%
\begin{equation}
S(\tsum_{i}p_{i}\rho _{i})\geq \tsum_{i}p_{i}S(\rho _{i})\text{.}  \label{20}
\end{equation}%
It means that our uncertainty about this mixture of states should be higher
than the average uncertainty of the states $\rho _{i}$.

The maximum amount of information that we can obtain about the identity of a
state is called the \textit{accessible information}. It is no greater than
the von Neumann entropy of the ensemble's density matrix (Holevo's theorem) 
\cite{21,22,23,24} and its greatest lower bound is the subentropy $Q(\rho )$ 
\cite{25} defined by

\begin{equation}
Q(\rho )=-\tsum_{j=1}^{n}\left( \tprod_{k\neq j}\frac{\lambda _{j}}{\lambda
_{j}-\lambda _{k}}\right) \lambda _{j}\ln \lambda _{j}\text{.}  \label{21}
\end{equation}%
The upper bound $\chi =S(\rho )-\tsum_{i}p_{i}S(\rho _{i})$, called
Holevo's, on the mutual information resulting from the measurement of any
observable, including POVM's, which may have more outcomes than the
dimensionality of the system being measured is%
\begin{gather}
H(A:B)\leq S(\rho )-\tsum_{i}p_{i}S(\rho _{i})\leq H(A)\text{,}  \label{22}
\\
H(A:B)\leq S(\rho )\leq \ln n\text{.}  \label{23}
\end{gather}

By analogy with the Shannon entropies it is possible to define conditional,
mutual and relative entropies%
\begin{gather}
S(A\mid B)\equiv S(A,B)-S(B)\text{,}  \label{24} \\
S(A:B)\equiv S(A)+S(B)-S(A,B)\text{,}  \label{25} \\
S(A:B)=S(A)-S(A|B)\text{,}  \label{26} \\
S(A:B)=S(B)-S(B|A)\text{.}  \label{27}
\end{gather}%
The negativity of the conditional entropy always indicates that two systems
are entangled and indeed, how negative the conditional entropy is provides a
lower bound on how entangled the two systems are \cite{6}.

The von Neumann entropy is additive, it means that if $\rho _{A}$ is a state
of system $A$ and $\rho _{B}$ a state of system $B$, then%
\begin{equation}
S(A\otimes B)=S(A)+S(B)  \label{28}
\end{equation}%
and strongly subadditive, which means that for a tripartite system in the
state $\rho _{ABC}$%
\begin{equation}
S(A,B,C)+S(B)\leq S(A,B)+S(B,C)\text{.}  \label{29}
\end{equation}%
Suppose distinct quantum systems $A$ and $B$ have a joint state $\rho _{AB}$%
. The joint entropy for the two systems satisfies the next inequalities%
\begin{gather}
S(A,B)\leq S(A)+S(B)\text{,}  \label{30} \\
S(A,B)\geq |S(A)-S(B)|\text{.}  \label{31}
\end{gather}%
The first inequality is known as subadditivity, and it means that there can
be more predictability in the whole than in the sum of the parts. The second
inequality is known as triangle inequality.

Let $\rho $ and $\sigma $ be density operators. We define the \textit{%
relative entropy }\cite{26}\textit{\ }of $\rho $ with respect to $\sigma $
to be%
\begin{equation}
S(\rho \parallel \sigma )=Tr(\rho \ln \rho )-Tr(\rho \ln \sigma )\text{.}
\label{32}
\end{equation}%
This function has a number of useful properties \cite{19}:

\begin{enumerate}
\item[1.] $S(\rho \parallel \sigma )\geq 0$, with equality if and only if $%
\rho =\sigma $.

\item[2.] $S(\rho \parallel \sigma )<\infty $, if and only if supp$\rho
\subseteq $ supp$\sigma $. (Here \textquotedblleft supp\textquotedblright\
is the subspace spanned by eigenvectors of $\rho $ with non-zero
eigenvalues).

\item[3.] The relative entropy is continuous where it is not infinite.

\item[4.] The relative entropy is jointly convex in its arguments \cite{27}.
That is, if $\rho _{1,}\rho _{2},\sigma _{1}$ and $\sigma _{2}$ are density
operators, and $p_{1}$ and $p_{2}$ are non-negative numbers that sum to
unity (i.e., probabilities), then 
\begin{equation}
S(\rho \parallel \sigma )\leq p_{1}S(\rho _{1}\parallel \sigma
_{1})+p_{2}S(\rho _{2}\parallel \sigma _{2})\text{,}  \label{33}
\end{equation}%
where $\rho =p_{1}\rho _{1}+p_{2}\rho _{2}$ and $\sigma =p_{1}\sigma
_{1}+p_{2}\sigma _{2}$. Joint convexity automatically implies convexity in
each argument, so that%
\begin{equation}
S(\rho \parallel \sigma )\leq p_{1}S(\rho _{1}\parallel \sigma )+p_{2}S(\rho
_{2}\parallel \sigma )\text{.}  \label{34}
\end{equation}
\end{enumerate}

Sanov's theorem \cite{3} has its quantum analogue \cite{28,29}. Suppose $%
\rho $ and $\sigma $ are two possible states of the quantum system $Q$, and
suppose we are provided with $N$ identically prepared copies of $Q$. A
measurement is made to determine whether the prepared state is $\rho $. The
probability $P_{N}$ that the state $\sigma $ is confused with $\rho $ is%
\begin{equation}
P_{N}\thickapprox e^{-NS(\rho \parallel \sigma )}\text{.}  \label{35}
\end{equation}%
The relative entropy can be seen as a measure of \ \textquotedblleft
distance\textquotedblright\ or of separation between of two density
operators. Two states are \textquotedblleft close\textquotedblright\ if they
are difficult to distinguish, but \textquotedblleft far
apart\textquotedblright\ if the probability of confusing them is small \cite%
{26}.

\end{document}